\documentclass[runningheads,a4paper]{llncs}

\usepackage{amssymb}
\usepackage{graphicx}

\usepackage{url}
\urldef{\mailsa}\path|{dzinoviev, zzhu, kjli}@suffolk.edu|    
\newcommand{\keywords}[1]{\par\addvspace\baselineskip
\noindent\keywordname\enspace\ignorespaces#1}

\begin{document}

\mainmatter  

\title{Building Mini-Categories in Product Networks}
\titlerunning{Building Mini-Categories in Product Networks}
\authorrunning{Building Mini-Categories in Product Networks}

\author{Dmitry Zinoviev\inst{1}%
\and Zhen Zhu\inst{2} \and Kate Li\inst{3}}
\institute{Department of Mathematics and Computer Science
\and Department of Marketing
\and Department of Information Systems and Operations Management\\
Suffolk University,\\
73 Tremont St., Boston, MA 02108, USA\\
\mailsa}

\maketitle

\begin{abstract}
We constructed a product network based on the sales data collected and
provided by a Fortune 500 Specialty Retailer. The structure of the
network is dominated by small isolated components, dense clique-based
communities, and sparse stars and linear chains and pendants. We used
the identified structural elements (tiles) to organize products into
mini-categories---compact collections of potentially complementary and
substitute items. The mini-categories extend the traditional hierarchy
of retail products (group--class--subcategory) and may serve as
building blocks towards exploration of consumer projects and long-term
customer behavior.

\keywords{retailing, product network, mini-category, category management}
\end{abstract}

\section{Introduction}

Consumer projects are large and major customer undertakings, often
involving a considerable amount of money, effort, and emotions. Examples
of consumer projects include porch renovation, Christmas decoration,
wedding planning, and moving into a college dorm. For each project,
customers often make multiple cross-category purchases through
multiple shopping trips. Such projects, in light of their significant
relevance to retailers' financial outcomes and customer
relationship~\cite{tuli2007}, are subject to thorough academic and
managerial investigations.

Theoretically, customer project management represents the frontier of
the category management domain, which is considered crucial by 72\% of
retailers surveyed by Kantar Retail in 2011. For years, most retailers
have been using only standard market research tools, mostly for
within-transaction product associations (e.g., market basket
analysis~\cite{agrawal1993}) and only from the functional or
manufacturers' perspectives for understanding product
categories~\cite{forte2013}. Few studies have explored product
association at the consumer project level.

The criticality of category management and the dearth of understanding
of consumers' project purchase behaviors serve as the impetus of this
research. This study aims to answer a key question: how to categorize
purchased products properly to prepare for project detection?
Equipped with the new advancements in complex network analysis
techniques~\cite{blondel2008,palla2005}, we expect our study to
discover product associations from the customers' view point, identify
mini-categories that serve as building blocks of project material
list, and provide guidance on managing project-level shopping
behaviors. In particular, we use Product Network Analysis (PNA) as the
primary analytical tool for this study. PNA applies Social Network
Analysis (SNA) algorithms to category management and is the automated
discovery of relations and key products within a product portfolio.

Methodologically, our research applies network analysis methods to
categorize products based on community discovery, a novel and
potentially insightful approach to the retailing field.  Managerially,
findings of this study will facilitate improving consumer-centric
category management beyond the traditional market basket
analysis~\cite{kim2012}. Our results will also provide guidance on
designing customized recommendation and promotion systems based on
identified project shopping behaviors~\cite{xie2014}.

The rest of the paper is organized as follows. We overview prior work
in Section~\ref{prior}. In Section~\ref{dataset}, we describe the data
set. In Section~\ref{network}, we explain the product network
construction algorithm. We explore the structure of the constructed
network and introduce mini-categories in Section~\ref{structure}. We
conclude and outline future work in Section~\ref{conclusion}.

\section{Prior Work\label{prior}}

Raeder and Chawla~\cite{raeder2011} are among the pioneers of product
network-leveled analysis. The authors follow an intuitive approach to
constructing a network of products from a list of sales transactions:
each node in the network represents a product, and two nodes are
connected by an edge if they have been bought together in a
transaction. Many real-world interaction networks contain communities,
which are groups of nodes that are heavily connected to each other, but
not much to the rest of the network. It is logical to expect that
product networks contain communities as well. Detecting communities in
complex networks is known as ``community
discovery''\cite{coscia2011}. In recent years, it has been one of the
most prolific sub-branches of complex network analysis, with dozens of
algorithms proposed and the agreement within the scientific community
that there is no unique solution to this problem given the many
different possible definitions of ``community'' for different
applications~\cite{pennacchioli2014}. Raeder and
Chawla~\cite{raeder2011} focus on community discovery in product
networks and show how communities of products can be used to gain
insight into customer behavior.

Pennacchioli et al.~\cite{pennacchioli2014} compare two community
discovery approaches: a partitioning approach, where each product
belongs to a single community, and an overlapping approach, where each
product may belong to multiple communities. The authors apply the
approaches to a data set of an Italian retailer and find that the
former is useful to improve product classification while the latter
can create a collection of different customer profiles. Xie et
al.~\cite{xie2014} provide a review and comparative study of
overlapping community discovery techniques. Videla-Cavieres and
R\'ios~\cite{videla2014} propose a community discovery approach based
on graph mining techniques that distinguishes two forms of
overlapping: crisp overlapping, where each product belongs to one or
more communities with equal strength; and fuzzy overlapping, where
each product may belong to more than one community but the strength of
its membership in each community may vary. Kim et al.~\cite{kim2012}
extend the idea of using only sales transaction data to build product
networks by utilizing customer information as well. The authors
construct two types of product networks: a market basket network
(MBN), which spatially expands the relationship between products
purchased together into relationship among all products using network
analysis; and a co-purchased product network (CPN), which is extracted
from customer-product bipartite network obtained using transaction
data. The topological characteristics and performances of the two
types of networks are compared.

\section{\label{dataset}The Data Set}
The data set provided to us through the Wharton Customer Analytics
Initiative (WCAI)~\cite{thd2014}, consists of product descriptions and
purchase descriptions.

The product part includes descriptions of ca. 111,000 material items,
351 non-material items (such as gift cards, warranties, deposits,
rental fees, and taxes), and 71 
items that combine materials and services. Since the descriptions of
the non-material items are generic and not easy to associate with
particular customer projects, we excluded them from our analysis.

The products are organized into a three-level hierarchy of 1,778
subcategories (e.g., \emph{landscaping}), 235 classes (e.g.,
\emph{live goods}), and 15 groups loosely corresponding to departments
(e.g., \emph{outdoor}). The members at each level in the hierarchy
are non-overlapping.

The purchase part contains the information of about 11,631,000
sales\footnote{For the purpose of this study, all items with the same
  product ID, purchased by the same customer at the same register at
  the same time, are considered one sale.} and 545,000 returns. For
each sale and return, we know the product ID, the buyer ID, and the
location (store ID and register ID), date, time, quantity, and price
of the sale, and discounts, if applicable. The sales recorded in the
data set took place over two years between 05/03/2012 and
02/03/2014. 99.6\% of the sales were initiated and completed in
stores; the remaining sales were made online.

The members at each level in the product hierarchy significantly vary
in size. The variance can be estimated in terms of the observed
entropy $H_1$ versus the entropy $H_0$ of a uniform, homogeneous
distribution of member sizes (higher entropy means higher
homogeneity). The data set group sizes range from 6 to 25,888
($H_1$=3.57 vs. $H_0$=3.91); class sizes---from 1 to 21,167 ($H_1$=4.37
vs. $H_0$=7.88); subcategory sizes---from 1 to 12,355 ($H_1$=4.55 vs.
$H_0$=10.79). The striking heterogeneity of the hierarchy members
makes it hard to treat them as first-order building blocks for further
research.

The data set product hierarchy reflects the store organization by
departments, sections, and subsections/shelves. While this grouping
makes perfect sense from the functional perspective (items performing
similar functions or intended for similar purposes, such as nails and
screws, are shelved together), it does not reveal latent task-oriented
connections between products. For example, 91\% of \emph{screws} are
in the \emph{hardware} and \emph{electrical} groups, but 82\% of
\emph{screwdrivers} are in the \emph{tools} group, another 18\% are in
the \emph{electrical} group, and none are in the \emph{hardware}. The
assignment of \emph{screws} and \emph{screwdrivers} to different
groups (and, therefore, different departments) ignores the fact that
both are required for \emph{screwdriving}. As a consequence, by
observing the purchase of \emph{screws} as an item from the
\emph{hardware} group and a \emph{screwdriver} as an item from the
\emph{tools} group, a researcher may not be able to detect that the
customer is about to start a \emph{screwdriving} ``project.''

To circumvent the problems of heterogeneity and lack of support for
task- or project-orientated classification, we introduce another level
in the data set hierarchy---mini-categories. We later define the
mini-categories as structural subnetworks within the overall product
network. The product network construction algorithm is described in
the next Section.

\section{\label{network}Product Network Construction}
A product network~\cite{forte2013,kim2012,pennacchioli2014} is a graph
$G$ reflecting the product co-occurrences in a customer's
``basket''~\cite{kim2012,raeder2011,videla2014}. The graph nodes
represent individual material items purchased by customers. Two nodes
$A$ and $B$ are connected with an edge if the products $A$ and $B$ are
frequently purchased together (not necessarily by the same
customer). The existence of an edge between two products suggests a
purposeful connection between the products, such as co-suitability for
a certain task, as in the \emph{screws} and \emph{screwdriver} example
above.

A product network graph $G$ is undirected (if $A$ is connected to $B$
then $B$ is connected to $A$). It does not contain loops (a node
cannot be connected to itself) or parallel edges ($A$ can be connected
to $B$ at most once). The graph in general is disconnected---it
consist of multiple components, one of which, the \emph{giant
  connected component} (\emph{GCC}), may have a substantially bigger
size than the others.

Depending on the construction procedure, the graph $G$ can be
unweighted or weighted. In the former case, the existence of an edge
indicates that the strength of the connection between the two incident
nodes (e.g., the likelihood of the two items to be in the same
``basket'') is simply at or above certain threshold $T$. In the latter
case, the strength of the connection is treated as an attribute of the
edge; this way, some edges are stronger than others. A weighted graph
can be converted to an unweighted graph by eliminating weak edges and
treating strong edges as unweighted. An unweighted, undirected graph
with no loops and parallel edges is called a simple graph.

While weighted graphs are more detailed, simple graphs are easier to
visualize and comprehend. Many graph processing algorithms (and
applications) are optimized for simple graphs. In our quest for
mini-categories, which are ambiguously defined, we believe that the
benefits of having a more detailed representation of product
interconnections are offset by the fuzzy mini-category detection
techniques, and do not outweigh the added complexity of handling
weighted graphs. That is why we chose simple graphs as the
representation of the product network.

At the first stage of the network construction, we create a graph node
for each material item that has been purchased by a customer
at least once over the observation period, to the total of 85,865
nodes.

\begin{figure}
\centering
\includegraphics[width=0.8\textwidth]{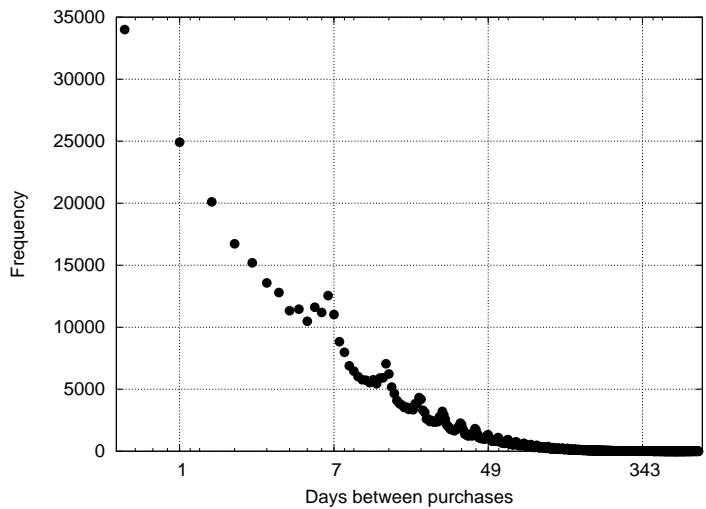}
\caption{Days between consecutive purchases by the same customer.}
\label{fig:intervals}
\end{figure}

At the second stage, two nodes are connected if the corresponding
items have been purchased \emph{together} at least $N$ times. To
quantify the concept of \emph{togetherness}, we first observed that
the customers are more likely to visit the store every
$k=1,2,3\ldots$ weeks (Fig.~\ref{fig:intervals}), which must be caused
by the weekly work cycle. We use one week as a natural window span and
consider two purchases by the same customer to be in the same
``basket'' if they were made within seven days (not necessarily within
one calendar week).

The choice of $N$ controls the density and the connectedness of the
product network. A bigger $N$ results in a sparse network with many tiny
isolated components that cannot be efficiently grouped into
mini-categories. A smaller $N$ yields a very dense network, dominated by
the GCC and unsuitable for community detection algorithms.

Table~\ref{tab:metrics} presents product network statistics for $N$=1,
5, 10, and 20: numbers of edges, nodes, isolated single nodes,
isolated pairs of nodes, and components; the size of the giant
connected component, and the relative volume of sales of the GCC
items. The two least dense networks ($N$=10 and 20) have a subtle
GCC and many isolates. The densest network ($N$=1) essentially
consists only of a very dense, nonclusterable GCC. The transition
from $N$=1 to $N$=5 substantially reduces the GCC size while
preserving its relative sales volume, thus making it possible, without
the loss of generality, to disregard the sales of the isolated
items. For this reason, we adopted $N$=5.

\begin{table}
\caption{Product network statistics for $N$=1, 5, 10, 20. See
  Section~\ref{staples} for the explanation of 5$^*$.}
\label{tab:metrics}
\centering
\begin{tabular}{lrrrrrr}\hline\noalign{\smallskip}
N & 1\phantom{.9\%} & 5\phantom{.9\%} & 5$^*$\phantom{.9\%} & 10\phantom{.9\%} & 20\phantom{.9\%}\\
\noalign{\smallskip}\hline\noalign{\smallskip}
Edges & 8,066,192\phantom{.9\%} & 104,643\phantom{.9\%} & 28,760\phantom{.9\%} & 26,187\phantom{.9\%} & 7,126\phantom{.9\%}\\
Nodes & 85,865\phantom{.9\%} & 85,865\phantom{.9\%} & 85,053\phantom{.9\%} & 85,865\phantom{.9\%} & 85,865\phantom{.9\%}\\
Isolated nodes & 1,026\phantom{.9\%} & 67,007\phantom{.9\%} & 69,619\phantom{.9\%} & 78,283\phantom{.9\%} & 82,982\phantom{.9\%}\\
Isolated pairs & 71\phantom{.9\%} & 682\phantom{.9\%} & 953\phantom{.9\%} & 494\phantom{.9\%} & 244\phantom{.9\%}\\
Components & 1,107\phantom{.9\%} & 67,989\phantom{.9\%} & 71,069\phantom{.9\%} & 79,051\phantom{.9\%} & 83,352\phantom{.9\%}\\
Absolute GCC size & 84,669\phantom{.9\%} & 16,215\phantom{.9\%} & 11,164\phantom{.9\%} & 5,296\phantom{.9\%} & 1,677\phantom{.9\%}\\
Relative GCC size & 98.6\% & 18.9\% & 13.1\% & 6.2\% & 2.0\%\\
Sales in the GCC & 99.9\% & 70.0\% & 51.3\% & 45.0\% & 26.0\%\\
\hline
\end{tabular}
\end{table}

\subsection{\label{staples}Staples}

\begin{figure}
\centering
\includegraphics[width=0.8\textwidth]{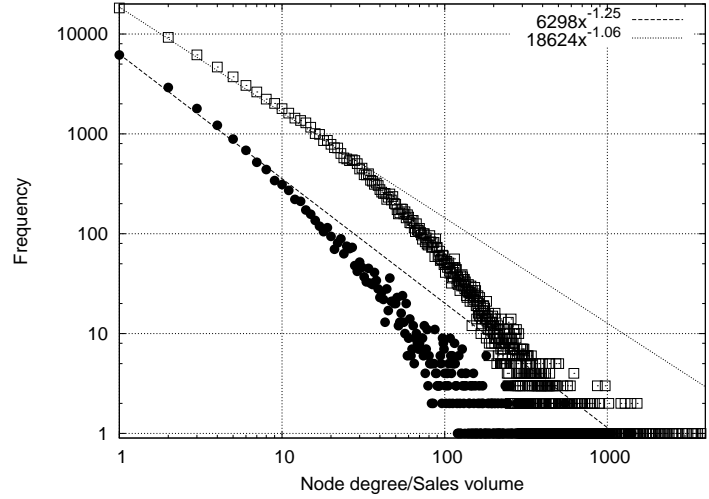}
\caption{Node degrees (\emph{circles}) and item sales volumes
  (\emph{boxes}) in the product network for $N$=5.}
\label{fig:degree}
\end{figure}

The resulting product network is a power-law graph with a long-tail
degree distribution with $\alpha\!\approx\!-1.25$
(Fig.~\ref{fig:degree}). The distribution of sales volumes for
individual items also follows the power law\footnote{In fact, node
  degrees and the corresponding sales volumes are correlated with
  $\rho\approx0.867$.} with $\alpha\!\approx\!-1.06$. Most items are
isolated nodes or have fewer than 10 connections. However, there is a
number of staples~\cite{brijs2004} in the tails of the distributions that are (a)
frequently purchased from the store and (b) frequently purchased
together with other items.

The top 20 staples in the product network are shown in
Table~\ref{tab:staples}.

The staples are either not related to any specific projects or are
generic and can be related to a multitude of projects. Since staples
belong to many ``baskets,'' they lay on many network shortest paths
and connect nodes that otherwise would probably be disconnected. The
shortest paths induced by the staples, increase graph coupling and
lower its modularity, thus eroding potential mini-communities. To
minimize the influence of the staples, we eliminate, in the spirit of
market basket analysis, 5\% of the GCC nodes with the highest
degrees---that is, 812 nodes with the degree $d>45$. The product
network $G^*$ with the truncated tail is referenced in
Table~\ref{tab:metrics} as 5$^*$.

\begin{table}
\caption{Top 20 most connected products (staples).}
\label{tab:staples}
\centering
\begin{tabular}{lrr}\hline\noalign{\smallskip}
Product & Degree & Sales\\
\noalign{\smallskip}\hline\noalign{\smallskip}
wood stud    &          1,410 & 3,305 \\
bucket                    &                  1,333 & 3,344 \\
plastic tray liner      &                  1,049 & 3,078 \\
biodegradable pot &                  1,031 & 3,756 \\
seal tape        &           986 & 2,491 \\
carbonated drink              &                    943 & 3,407 \\
adhesive tape      &                   897 & 2,258 \\
diet soda                      &                    810 & 2,897 \\
flat brush                  &                    715 & 2,241 \\
drywall    &     681 & 1,395 \\
tray set       &                    677 & 2,051 \\
topsoil                        &                    674 & 3,498 \\
vegetable peat pot       &                    634 & 2,706 \\
insulating foam sealant       &                   613 & 2,103 \\
plastic bag                    &                    593 & 2,529 \\
latex caulk                      &                    587 & 1,453 \\
garden soil  &                   586 & 2,753 \\
drinking water                    &                    556 & 2,523 \\
poly roll &             549 & 1,524 \\
contractor trashbag  &                    549 & 2,057 \\
\hline
\end{tabular}
\end{table}

\section{\label{structure}Network Structure and Mini-Categories}

A visual inspection of $G^*$ reveals rich internal structure of the
product network. In particular, we noticed three major types of
structural tiles: dense clique-based communities, sparse stars, and
linear chains and pendants---and randomly structured connecting
matter. Often, the tiles overlap (e.g., a node can be a leaf of a star
and a member of a dense community). We propose an automated procedure
for the structural tile extraction.

\subsection{Tile Extraction}
At the pre-processing stage, all small unconnected components (having
fewer than five nodes) are removed from the network. The new network
has 12,416 nodes and 26,943 edges.

We define an imperfect star as a connected subgraph of $G^*$ that
consists of at least four nodes of degree $\le2$, connected to a
common central node. We allow for a modest number ($n/2$) of chords in
an $n$-node star, because the graph $G^*$ was constructed through a
binarization procedure with an arbitrary chosen threshold and an
absence of a connection between two nodes does not imply a zero
co-occurrence. 

A chain/pendant is a linear sequence of nodes that is connected to
anchor nodes at one (pendant) or both (chain) ends. We define an
imperfect chain/pendant (a linear tile) as a connected subgraph of $G^*$
that consists of nodes of degree 1 through 3. The nodes of degree 3
introduce defects (chords and mini-stars) but do not significantly
distort the linear structure of the subgraphs.

An anchor node is a node that is shared by a linear tile and the rest
of $G^*$. We attached anchor nodes to the incident linear tiles. As a
result, we get 5,197 small linear tiles with $<5$ nodes and 375 large
linear tiles with $\ge5$ nodes. In the spirit of restricting the size
of individual tiles to $\ge5$ nodes, we combined the small linear
tiles with their larger immediate neighbors.

We used CFinder~\cite{palla2005} for the extraction of dense
communities. CFinder is based on the Clique Percolation Method: it
builds $k$-cliques---fully connected subgraphs of $G^*$ of size
$k$---and then computes the union of all $k$-cliques that share $k-1$
nodes pairwise. Clique-based communities have an important advantage
over $k$-cliques: they are less rigidly defined and can absorb more
potentially related nodes than a clique, thus improving the tile
coverage of $G^*$ and reducing the number of required tiles.

We eliminated communities with $<5$ nodes to be consistent with the
previously adopted approach to small tiles.

\subsection{Coverage Optimization}
As a result of the network decomposition, we constructed 5,035
possibly overlapping tiles of three different types: stars (3,553),
dense clique-based communities (1,107), and chains/pendants (375). The
union of all tiles contains 12,370 product network nodes, with the
average coverage of 2.45 nodes per tile. Table~\ref{tab:tiles} shows
the summary of the tile coverage (before and after optimization).

\begin{table}
\caption{Structural tiles of the product network before and after
  coverage optimization.}
\label{tab:tiles}
\centering
\begin{tabular}{lrrrrr}\hline\noalign{\smallskip}
Tile type & \multicolumn{2}{c}{Count} & \multicolumn{2}{c}{Node Coverage}&Mean Size\\
\noalign{\smallskip}\cline{2-6}\noalign{\smallskip}
& Original & Optimized & Original & Optimized & Optimized\\
\noalign{\smallskip}\hline\noalign{\smallskip}
Stars & 3,553 & 289 & 10,486 & 5,589&30\\
Dense communities & 1,107 & 216& 5,457 & 8,123&47\\
Pendants/chains & 375 & 313& 2,065 & 4,278&17\\
\noalign{\smallskip}\hline\noalign{\smallskip}
Total: & 5,035 & 818 & 12,370 & 12,274&\\
\hline
\end{tabular}
\end{table}

The amount of overcoverage (average number of tiles that a node
belongs to) can be reduced by optimizing the coverage, identifying
essential tiles, and discarding redundant tiles. For optimization, we
chose a variant of a greedy maximum coverage
algorithm~\cite{johnson1974}. We start with an empty set of covering
tiles. At each iteration, we select an unused tile that, if added to
the coverage set, minimizes the number of uncovered nodes and
increases the number of covered nodes. The process stops when no such
tile exists.

\begin{figure}
\centering
\includegraphics[width=0.65\textwidth]{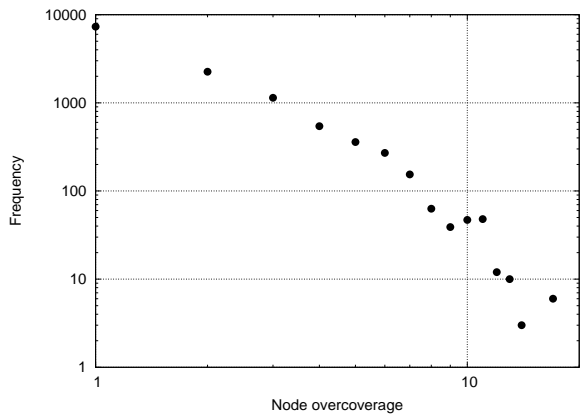}
\caption{Overcoverage (number of structural tiles per node) of product
  network nodes.}
\label{fig:overcoverage}
\end{figure}

The optimization reduced the number of essential tiles to 818---16\%
of the original tile set (see Table~\ref{tab:tiles}). Only 142 nodes
remained uncovered by any essential tile. As a result, the average
number of nodes per tile increased to 15, and the amount of
overcoverage was dramatically reduced (Fig.~\ref{fig:overcoverage}).

\begin{figure}
\centering
\includegraphics[height=3.cm]{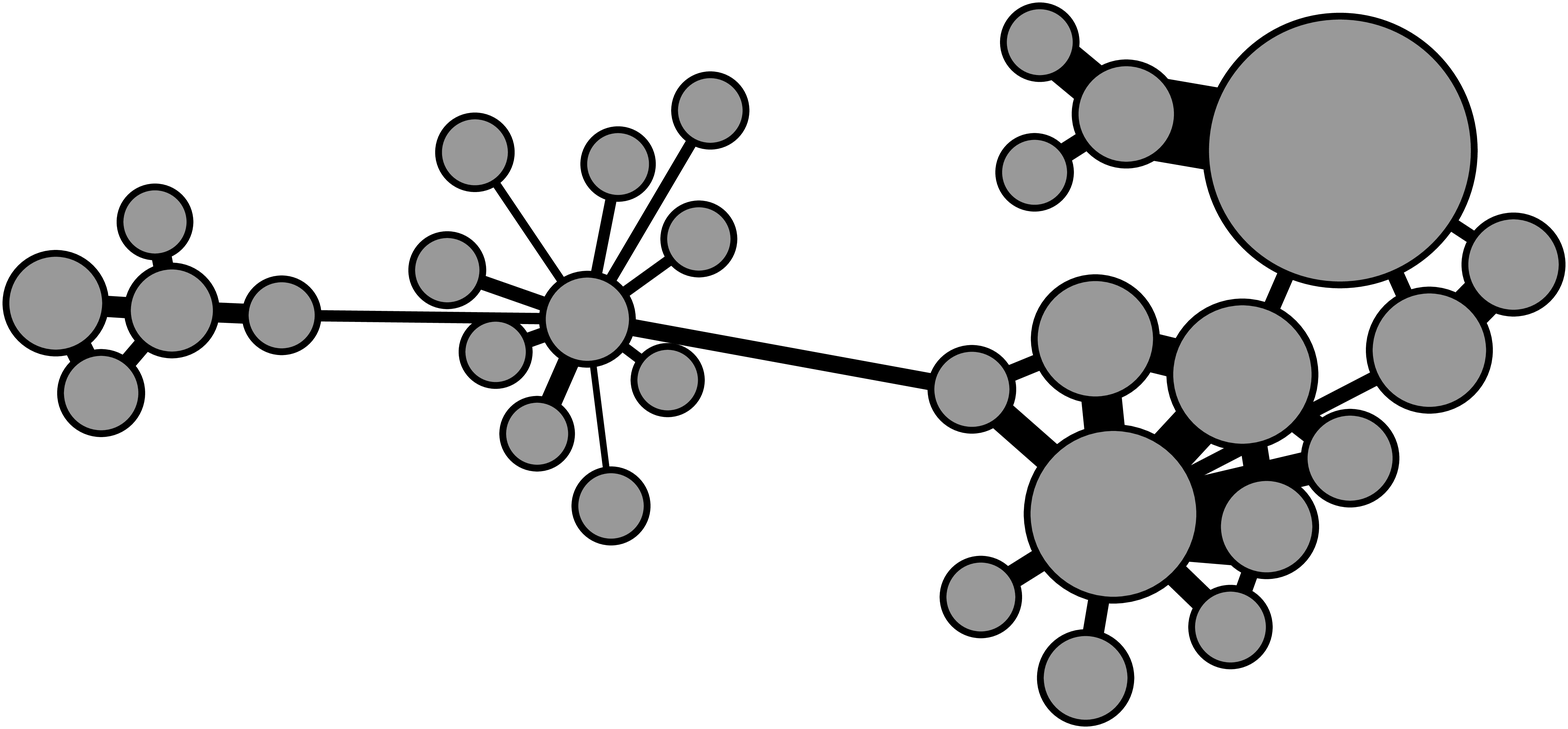}
\hspace{5mm}
\includegraphics[height=4.cm]{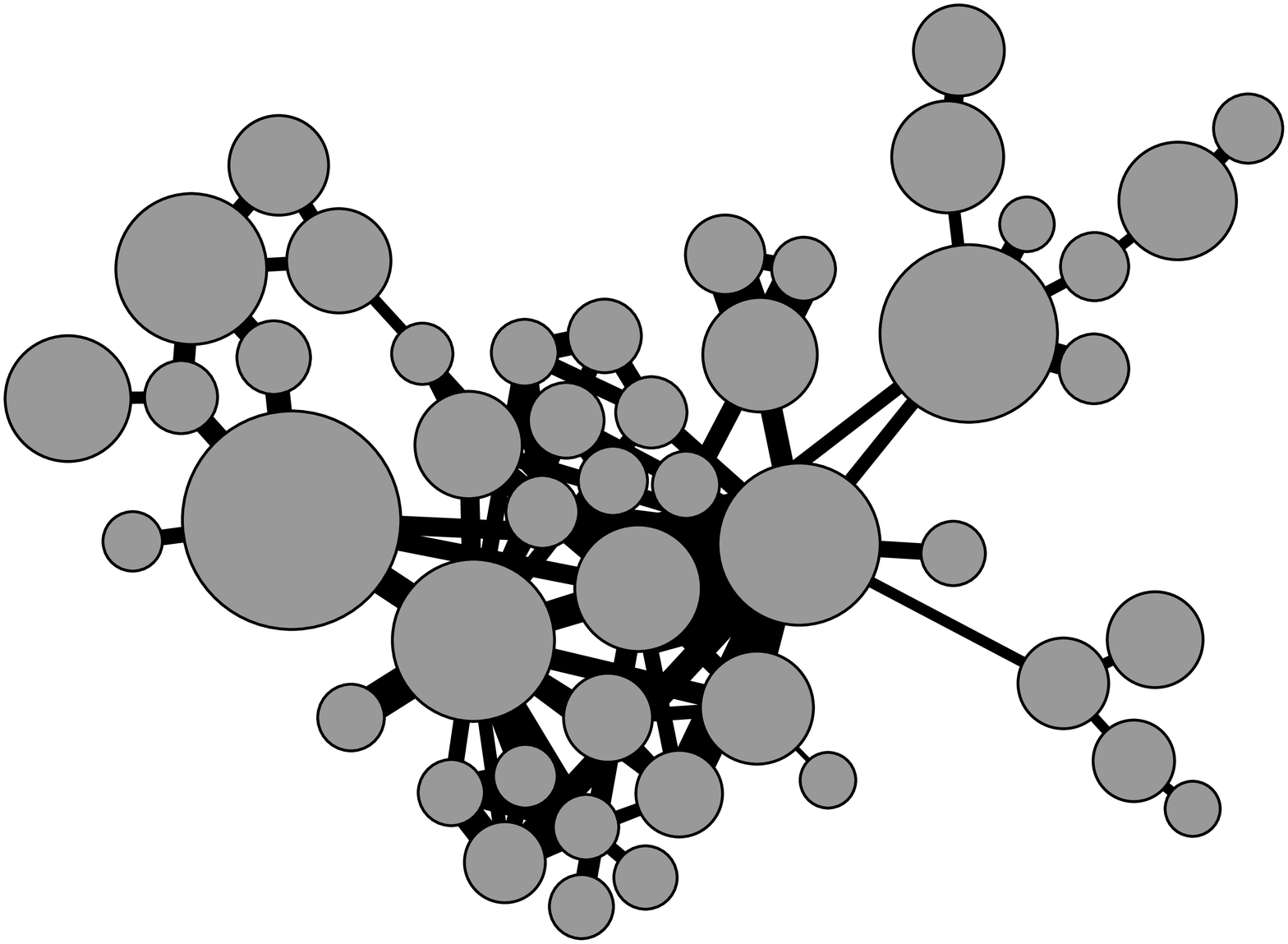}
\hspace{5mm}
\includegraphics[height=2.cm]{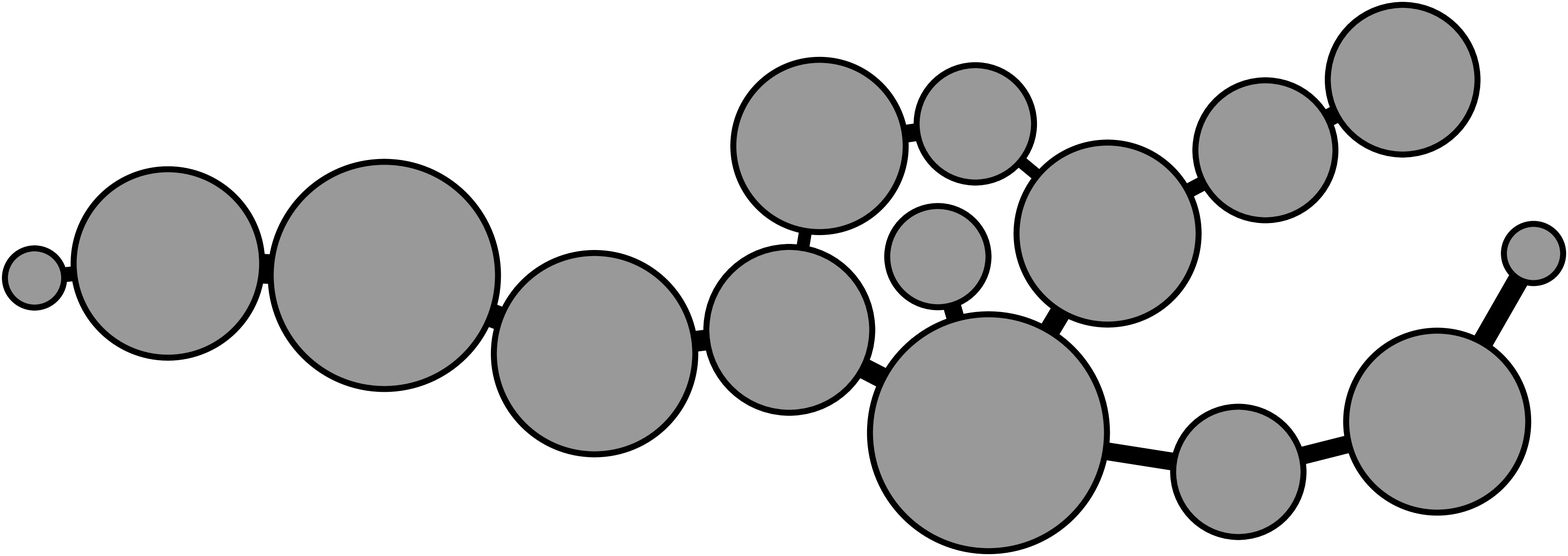}
\caption{Outlines of sample tiles: (a) star ``\emph{Ice
    melt and shovels}'' (\emph{top}), (b) community ``\emph{Alarms
    and smoke detectors}'' (\emph{bottom left}), and (c) chain
  ``\emph{Zinc screws}'' (\emph{bottom right}). Node size represents
  item sale volume, edge thickness---the number of co-occurrences.}
\label{fig:samples}
\end{figure}

Figure~\ref{fig:samples} shows the outlines of three randomly selected
average-sized sample tiles of each type.

\subsection{Mini-Categories}

The optimized tile set contains a reasonable number of members and has
a good uniformity. The entropy of the tile size distribution for the
set is $H_1\approx8.92$ versus $H_0\approx9.68$ for the uniform,
homogeneous distribution. The collection of essential tiles forms a
good structural basis for further research of customer behavior and
customer-driven projects.

From a retailing perspective, different types of structural tiles
correspond to different relationships between the products associated
with the tile nodes. We call these building blocks mini-communities
and suggest that they reflect consumers' view on the product
hierarchy.

A cliques-based community (and especially a generating $k$-clique) is
characterized by homogeneity and complete or almost complete
connectivity between the nodes. In other words, any product in a
community is commonly purchased together with all other products in
the community. The products in a community form a topical
complementary group~\cite{brijs2004,lattin1985,elrod2002}, e.g., \emph{alarms}
and \emph{smoke detectors}---elements of home security.

On the contrary, a star is heterogeneous. The nodes in a star form two
different groups: the high-degree hub (the lead product) and
small-degree spokes (the peripheral products). The lead product is
frequently purchased together with one or few peripheral
products. However, the peripheral products are never or almost never
purchased together. The hub with the peripherals form a group of
substitutes~\cite{brijs2004,lattin1985,elrod2002}, e.g., snow removal
tools and materials: \emph{ice melt bag} as the lead and
\emph{shovels}, \emph{rock salt}, and \emph{sand} as the peripherals
(Figure~\ref{fig:samples}a).

Chains/pendants (linear tiles) are perhaps the hardest mini-category
to interpret. They describe a set of products that are almost never
purchased together, but often purchased pairwise. An almost perfect
example of a chain is shown in (Figure~\ref{fig:samples}c): all
products in the tile are \emph{zinc wood screws} and
differ only in length and number (diameter). Most of the screws are
\#8 and \#10. Any two neighbors differ either in diameter (\#8 vs.
\#10) or length, and the difference between the neighbors is always
smaller than between any non-neighbors. We hypothesize that a customer
buys a pair of items if she is not sure about the precise values of
certain attributes (such as screw dimensions). In other words, a
linear tile represents substitutes by ignorance, as opposed to
substitutes by choice.

\section{\label{conclusion}Conclusion and Future Work}
The goal of this research is to pave the road to the automated
identification of consumer projects, based on the available retail
data. One possible direction that we explored is to deconstruct the
product network into structural tiles that correspond to groups of
products---mini-categories. 

We built a product network from the purchase data provided by a
Fortune 500 Specialty Retailer through the Wharton Customer Analytics
Initiative (WCAI). A visual inspection of the network revealed three
major types of structural blocks: dense clique-based communities,
stars, and linear structures (chains and pendants).

We developed a procedure for the automated tile extraction and
coverage optimization. As a result, we produced a reasonably uniform
in size collection of ca.~800 tiles of all three types that cover the
majority of the giant connected component of the product network. We
associate each tile type with the nature of the products in the tile:
either complements or substitutes.

We believe that the extracted mini-categories represent consumer view
on the retail product hierarchy and can be used as an efficient
managerial and research tool.

In the future, we plan to study mini-categories as first-class
objects, rather than building blocks for possible consumer
projects. That way, there will be no need to minimize their count and
lump mini-chains into adjacent stars and cliques, thus preserving
relative cleanness of the stars and cliques and making them easier to
analyze.

We hope that the planned use of structural role extraction
algorithms~\cite{henderson2012} will uncover more tile categories,
that, in turn, would yield more retailing-related mini-categories.

Finally, we will look into validating our complement/substitute tile
theory using Amazon Mechanical Turk~\cite{buhrmester2011}
crowdsourcing platform.

\subsubsection*{Acknowledgments.} 
The authors would like to thank Wharton Customer Analytics Initiative
(WCAI) for the provided data set that made this research possible and
an anonymous reviewer for the suggestion to use role extraction
algorithms.

\end{document}